\documentclass[%
 aip,
 amsmath,amssymb,
 reprint,%
floatfix]{revtex4-2}

\usepackage{graphicx}
\usepackage{dcolumn}
\usepackage{bm}

\usepackage[utf8]{inputenc}
\usepackage[T1]{fontenc}
\usepackage{mathptmx}
\usepackage{etoolbox}
\usepackage{cleveref}
\usepackage{xcolor}
\usepackage{subcaption}
\usepackage{booktabs}
\usepackage{siunitx}
\usepackage{makecell}

\newcommand{\NbOx}{\ensuremath{\mathrm{Nb_2O_5}}}
\newcommand{\HF}{\ensuremath{\mathrm{H_2/SF_6}}}

\makeatletter
\def\@email#1#2{%
 \endgroup
 \patchcmd{\titleblock@produce}
  {\frontmatter@RRAPformat}
  {\frontmatter@RRAPformat{\produce@RRAP{*#1\href{mailto:#2}{#2}}}\frontmatter@RRAPformat}
  {}{}
}%
\makeatother

\begin{document}

\preprint{AIP/123-QED}

\title{Atomic layer etching of niobium via alternating O$_2$ and H$_2$/SF$_6$ plasma exposures}

\author{Lillian B. Hughes Wyatt}
 \affiliation{Division of Engineering and Applied Science, California Institute of Technology, Pasadena, CA 91125, U.S.A.}\

 \author{Azmain A. Hossain}
 \affiliation{Jet Propulsion Laboratory, California Institute of Technology, Pasadena, CA 91109, U.S.A.}

\author{Yifei Yan}
\affiliation{Division of Engineering and Applied Science, California Institute of Technology, Pasadena, CA 91125, U.S.A.}

\author{Austin J. Minnich}
\affiliation{Division of Engineering and Applied Science, California Institute of Technology, Pasadena, CA 91125, U.S.A.}
 \email{aminnich@caltech.edu}

\date{\today}

\begin{abstract}
Niobium thin films are a key component of superconducting detectors and electronics, and film surface quality, surface roughness, and geometrical dimensional accuracy affect many performance metrics. 
Atomic layer etching (ALE) is of interest for modifying film surfaces owing to its Angstrom-scale etching precision and surface smoothing characteristics.  
Here, we report a plasma-enhanced ALE process for Nb based on sequential O$_2$ plasma and \HF{} plasma exposures. 
The process achieves an etch per cycle of 5--7 \AA/cycle and exhibits surface smoothing.
The superconducting transition temperature of the ALE-treated films is maintained up to the expected decrease associated with reduced film thickness. 
Our ALE process could find applications in superconducting electronics for which surface quality and precise thickness control are of high importance.
\end{abstract}

\maketitle

\section{Introduction}
Niobium thin films are a versatile platform for superconducting devices, combining a relatively high critical temperature with well-established deposition and microfabrication processes.
These characteristics have facilitated their use in a wide range of applications from superconducting radio-frequency cavities\cite{Gurevich2017Theory} and microwave resonators\cite{Zmuidzinas2012Superconducting, wallraffStrong2004} to single photon detectors\cite{Semenov2003Ultimate} and superconducting quantum circuits.\cite{SiddiqiEngineeringHighcoherence2021c, anferovImproved2024}
Device fabrication often requires precise pattern transfer and control of film thickness to achieve the desired device performance.\cite{drimmerEffect2025, pintoDimensional2018}
As a consequence, the etching process must accurately transfer the geometrical pattern from the mask while minimizing surface damage and roughness and preserving the superconducting properties of the underlying film.\cite{murrayMaterial2021, altoeLocalization2022a, deLeonMaterials2021}

Reactive ion etching (RIE) and other conventional plasma etching methods are widely used for Nb etching.\cite{lichtenbergerInvestigation1993, anferovImproved2024, drimmerEffect2025}
However, RIE is a continuous process, and achieving nanometer-scale etch depth control is challenging. Further, energetic ion bombardment can introduce film damage and redeposition that cause microwave loss and other undesirable effects. \cite{anferovImproved2024, kanarikOverview2015, murrayMaterial2021}
Atomic layer etching (ALE) provides an alternative approach to etching in which the etching process is separated into sequential surface modification and removal steps.  \cite{kanarikOverview2015}
In an ideal ALE process, each step is self-limiting such that a certain thickness of material is modified and subsequently removed during each cycle regardless of the amount of reactant supplied, and the total etch depth is controlled through the number of cycles. 
ALE has been shown to achieve Angstrom-scale etch per cycle while also smoothing surfaces to sub-nanometer roughness. \cite{kanarikAtomic2018}

Recently, surface treatments combining ALE removal of native oxides  with ALD encapsulation as the final processing step have demonstrated reduced dielectric losses in Al-based superconducting quantum devices.\cite{mahuliImproving2025} However, ALE processes for superconducting Nb have received limited attention. Nb ALE has been only recently shown  using CF$_4$/H$_2$ plasma modification followed by Ar$^+$ ion irradiation.\cite{WalshAtomicLayer2026a} While this demonstration provides a foundation for ALE processing of Nb, the effects of the ALE process on the composition of the near-surface region and the resulting superconducting and electrical properties of the films have not yet been extensively investigated.

Here, we report a plasma-enhanced ALE process for Nb using sequential O$_2$ plasma and \HF{} plasma exposures.
We characterize the etch per cycle and its dependence  on the half-cycle conditions and examine the resulting surface morphology and chemical composition. 
The process exhibits surface smoothing while maintaining the superconducting transition temperature within the expected decrease associated with reduced film thickness.\cite{pintoDimensional2018, Yifei}
Our results may find applications in superconducting detectors and qubits which rely on precise control of Nb film thickness and surface quality.

\section{Methods}
The ALE process was performed in an Oxford Instruments FlexAL II atomic layer deposition (ALD) system using an inductively coupled plasma (ICP) source.
The table temperature was set to ${125}~^\circ\text{C}$.
Before ALE processing, the chamber and Al$_2$O$_3$-coated carrier wafer were conditioned with 10 minutes of SF$_6$/O$_2$ plasma and at least 300 cycles ($\sim50$ nm) of Al$_2$O$_3$ ALD, followed by a final 5 minutes of SF$_6$ plasma.
As in ref.~\cite{hossainAtomic2025}, alumina was chosen because it forms a stable AlF$_3$ layer after SF$_6$ plasma exposure.  

\begin{figure}
    \vspace{2 mm}
    \centering
    \includegraphics[width=1.0\linewidth]{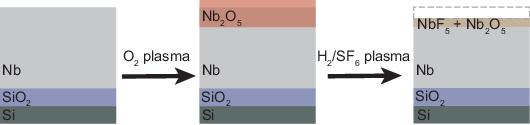}
    \caption{Schematic of the Nb ALE process. The Nb surface is first oxidized via O$_2$ plasma to create \NbOx{}. The oxide is then converted to NbF$_5$ and selectively etched using \HF{} plasma.}
    \label{fig:overview}
\end{figure}

Our ALE approach, depicted in \Cref{fig:overview}, consists of sequential oxidation and selective removal of the oxide by conversion to a high-volatility fluoride. 
It was inspired by prior ALE processes developed for superconducting metal nitrides such as NbN\cite{hossainAtomic2025} and TiN \cite{hossainIsotropic2023} which exploit the difference in oxidation state of the metal between the bulk film and the oxide to achieve an etching contrast. 
In the prior works on metal nitrides, oxidation with oxygen plasma is used to preferentially place the Nb in the +5 oxidation state, which then produces volatile NbF$_5$ and hence etching. \cite{hossainAtomic2025, hossainIsotropic2023}  
\HF{} plasma is used for the fluorination step because pure SF$_6$ spontaneously etches Nb. 
Instead, H$_2$ is added to the SF$_6$ plasma to promote HF molecule formation and suppress F radical production.\cite{hossainAtomic2025, krieger2025atomic}
Prior studies have demonstrated this shift in plasma chemistry at low SF$_6$:H$_2$ ratios,\cite{Pankratiev2020Etching, Jung2020Highly, krieger2025atomic} suggesting that H$_2$ dilution can be tuned to minimize spontaneous Nb etching while maintaining the desired \NbOx{} etch.

In more detail, the first half-step of the ALE process consists of a 10 s, 100 W O$_2$ plasma exposure with 50 sccm O$_2$ at 100 mTorr, oxidizing the Nb predominantly into the +5 oxidation state. \cite{hossainAtomic2025}
A plasma power of 100 W was selected as the lowest power that consistently sustains a stable plasma while minimizing potential plasma-induced damage.
After oxidation, the chamber is flushed with 500 sccm Ar for 15 s and pumped out for 10 s.
The second half-step etches the modified region of \NbOx{} by fluorination to NbF$_5$ via a 50 s, 100 W \HF{} plasma with 20 sccm H$_2$ and 7 sccm SF$_6$ at 100 mTorr. 
The 100 W power was again chosen to minimize plasma damage and to prevent spontaneous etching observed at higher \HF{} plasma powers up to 400 W.
The chamber is again flushed with 500 sccm Ar for 15 s and pumped out for 10 s to conclude a full ALE cycle.

Immediately before each etch rate measurement for a particular condition, a fresh niobium film was exposed to a 100 W SF$_6$ plasma for 7 s, followed by 10 cycles of ALE to remove the native oxide. This procedure was performed to produce a consistent starting surface condition across all samples by eliminating any effect of mixed oxidation states in the native Nb oxide surface.

Nb samples used in this work were deposited either by RF sputtering on SiO$_2$/Si substrates in an AJA ATC Orion 8 UHV sputtering tool or by thermal laser evaporation (TLE) on sapphire substrates in a home-built system.\cite{Yifei}
All samples in this work were cleaved to be 10x10 mm$^2$ in size.
Substrates for sputtering consisted of an 800 nm-thick thermal SiO$_2$ layer grown on Si, providing a dielectric layer beneath the sputtered metal film that enables thickness measurements by ellipsometry.
Before sputtering, the substrates were cleaned in solvents (sonication in acetone, IPA) and exposed to a 30 s, 100 W O$_2$ descum in a Tergeo Plus plasma cleaner.
After loading in the sputterer, the substrates were heated in vacuum at ${200}~^\circ\text{C}$ for 20 minutes to degas atmospherics and then Ar$^+$ ion milled for 3 minutes (30W substrate RF bias) while cooling down to the deposition temperature of ${100}~^\circ\text{C}$.
Nb was sputtered with 130 W / 10 sccm Ar gas flow / 2 mtorr / 30 W RF substrate bias, resulting in a deposition rate of 0.3 \AA/s.  
We note that this sputtering recipe was not optimized, which is beyond the scope of this work. 
Sputtered films were used for process optimization and surface analysis by XPS.
For electrical and superconducting properties and roughness characterization, we used the higher quality TLE-grown Nb films.
Details of the TLE growth procedure can be found in ref.~\cite{Yifei}

Nb film thickness was characterized ex-situ using a J. A. Woollam M-2000 spectroscopic ellipsometer by measuring a nine-point grid across the chip at incidence angles of ${60}^\circ$ and ${70}^\circ$ over a wavelength range of 370–1000 nm. 
The ellipsometry data were fit using a Cauchy/B-spline/SiO$_2$/Si multilayer model, with the Cauchy model describing the surface oxide and the B-spline model used to determine the Nb metal thickness. 
Measurements at nine locations across the chip enable the film thickness and its spatial variation to be quantified from the mean and standard deviation of the extracted thicknesses. 
The underlying dielectric layer enables reliable ellipsometry measurements of Nb films with thicknesses up to approximately 50 nm.

X-ray photoelectron spectroscopy (XPS) was performed using a Kratos Axis Ultra spectrometer equipped with a monochromatic Al K$\alpha$ X-ray source (1486.6 eV) and a base pressure of $ <3\times10^{-9}$ Torr. 
The hemispherical electron analyzer was oriented for detection along the sample surface normal. 
The instrument was operated using Vision Manager software v2.2.10 revision 5, and XPS data were analyzed using CasaXPS software v2.3.25PR1.0 (CASA Software Ltd.).
All peaks were fit in CasaXPS with a Shirley background subtraction, and Nb peaks were analyzed according to the binding energies categorized in the NIST XPS database.\cite{NISTXPS}
Nb doublets were constrained to have a spin–orbit splitting of 2.72 eV, an intensity ratio of 3:2 between $3d_{5/2}$ and $3d_{3/2}$, and identical FWHM.
The unoxidized, pure Nb peaks were fit with an asymmetric Lorentzian LA(0.9,1.8,12), as in ref.~\cite{Kar2023,Prudnikava2022Niobium}
Nb oxides with metallic conductivity (Nb$_2$O and NbO) were fit with LA(1.2,5,12), and non-metallic oxides (\NbOx{} and NbO$_2$ were fit with a 50$\%$ Lorentzian and 50$\%$ Gaussian (GL50) lineshape.\cite{Dhas2025,Kar2023} 
All peaks for other elements were fit with the CasaXPS standard default GL30 lineshape.

Cryogenic electrical measurements were performed on a Quantum Design DynaCool Physical Property Measurement System (PPMS) using the DC resistivity option. 
Resistivity ($\rho$) was measured on Nb films wirebonded with four Al wires to a DC resistance puck, first at 300 K using the Van der Pauw method and subsequently at cryogenic temperatures using a four-point configuration.
In the PPMS, the resistivity was measured over a temperature range of 2–300 K to determine the superconducting critical temperature ($T_c$) and residual resistance ratio (RRR), defined here as $\rho_{300\mathrm{K}}/\rho_{20\mathrm{K}}$.

\section{Results}
\subsection{Atomic layer etching of niobium thin films}

\begin{figure}
    \vspace{2 mm}
    \centering
    {\includegraphics[width=1.0\linewidth]{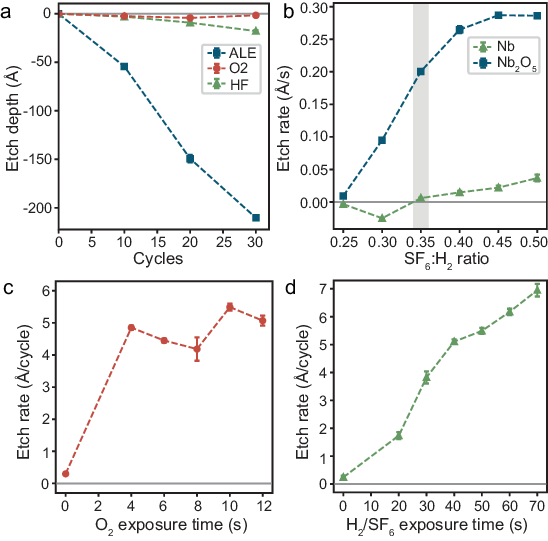}
    \phantomsubcaption\label{fig:etchdepth-cycles}
    \phantomsubcaption\label{fig:etchrate}
    \phantomsubcaption\label{fig:halfcycle1}
    \phantomsubcaption\label{fig:halfcycle2} }
    \caption{(a) Nb etch depth versus number of cycles for exposure only to O$_2$ (red circles) and \HF{} (green triangles) plasma half-steps and both half-steps (blue squares). (b) Etch rate of the Nb (green triangles) and intentionally oxidized Nb surface (blue squares) versus the SF$_6$:H$_2$ gas flow rate ratio. The shaded region highlights the flow rate ratio selected for the ALE process due to its high etch selectivity of \NbOx{} over Nb. (c) Nb etch per cycle (EPC) versus O$_2$ plasma exposure time with \HF{} plasma exposure time fixed at 50 s. (d) EPC versus \HF{} plasma exposure time, with the O$_2$ half-step fixed at 10 s. The EPC saturates with increasing exposure time for the O$_2$ half-step, while for \HF{} half-step, soft saturation occurs around 50 s before the etch rate continues to increase. In all plots a solid gray line at y=0 is plotted for reference and dashed lines serve as guides to the eye. Error bars represent the standard deviation of thickness measured over a nine-point grid on the chips.}
    \label{fig:ALE}
\end{figure}

\begin{figure*}
    \vspace{2 mm}
    \centering
    \includegraphics[width=1.0\linewidth]{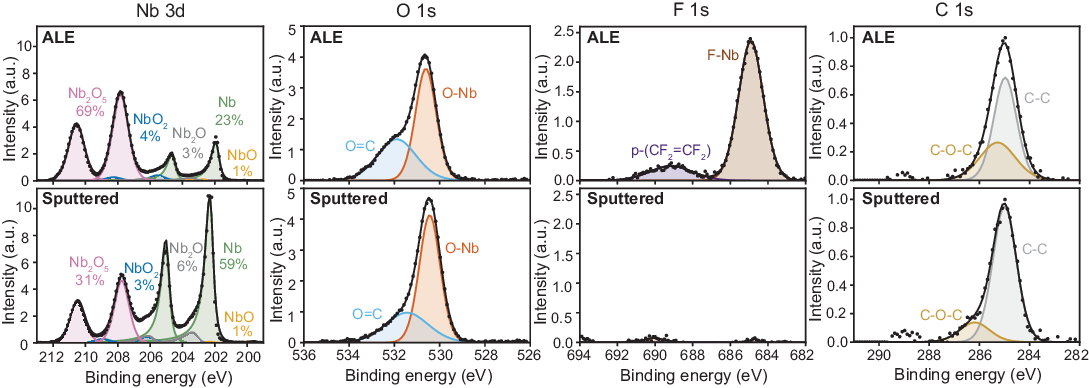}
    \caption{Surface XPS spectra for ALE and sputtered samples, showing (from left to right) Nb 3d, O 1s, F 1s, and C 1s peaks. The data are background-subtracted (Shirley) and normalized to the C 1s maximum for each respective sample. The black line shows the envelope of the fit, which is composed of underlying components, colored and labeled.}
    \label{fig:XPS}
\end{figure*}

We begin by showing the etching characteristics of the process. \Cref{fig:etchdepth-cycles} shows the Nb etch depth versus cycle number for each half-step and the full ALE process. 
The half-steps separately yield negligible etching, while the full cycle exhibits  an etch per cycle (EPC) of 5--7 \AA/cycle. 
The process synergy is around 90\%. 
Here, synergy was computed as the difference between the total etch per cycle and the sum of etching during individual half-cycles, divided by the total etch per cycle.
The small amount of \HF{} etching observed at increased cycles may be due to the conversion of residual \NbOx{} and/or lower oxidation state fluorides to NbF$_5$ over time. 
We note that the EPC is very sensitive to the chamber condition, and because our experiments were performed in a shared ALD tool we observe up to $\sim$ 2 \AA/cycle variation in the etch per cycle when other processes (particularly those involving Nb, Ti, or Si precursors) have recently been performed.

\Cref{fig:etchrate} shows the selective etching of plasma-oxidized \NbOx{} compared with pure Nb in \HF{} plasma with varying SF$_6$:H$_2$ gas flow ratios.
At SF$_6$:H$_2$ ratios $>$ 0.40, the substantial concentration of F radicals in the plasma produce spontaneous etching of Nb, while at lower ratios $<$ 0.30, negligible etching of either Nb or \NbOx{} occurs.
We chose the 0.35 SF$_6$:H$_2$ ratio (highlighted in gray) as the standard for our ALE process since it displays the best selectivity of \NbOx{} over Nb.

Lastly, \Cref{fig:halfcycle1,fig:halfcycle2} show saturation curves for each half-step, where the exposure time of one step is varied while the other is held constant.
The oxidation half-step exhibits self-limiting behavior, as the EPC plateaus at longer O$_2$ exposure times. 
We therefore select 10 s for the O$_2$ plasma half-step, which is well within the saturation regime.
The fluorination half-step exhibits less distinct saturation; the increase in EPC slows as the \HF{} exposure approaches 50 s but then increases further at longer exposure times.
We hypothesize that the initial saturating regime corresponds to conversion and removal of the O$_2$ plasma-modified \NbOx{} layer, while at longer \HF{} exposures, slower reaction with deeper \NbOx{} or suboxide species may drive further material removal and the second increase in EPC.
The best ellipsometry fit for the ALE-treated samples shows a thick Cauchy layer at the Nb surface ($\sim$ 4--6 nm), consistent with prior reports of increased \NbOx{} thickness after O$_2$ plasma exposure\cite{GiacconeStudyNiobium2022} and thicker than a typical native oxide ($\sim$ 2--4 nm).
The substantial thickness of the Cauchy layer supports the presence of an extended plasma-modified surface region beyond the native oxide, which may contribute to continued etching at long \HF{} exposures.
We selected the soft saturation regime with 50 s \HF{} exposure time for further studies since this time is sufficient to produce atomic layer etching and avoids technical issues related to difficulty of pumping HF byproducts with longer \HF{} plasma times in our tool.

\subsection{Surface composition and roughness}
We next performed XPS on both ALE-treated and as-sputtered films to understand the effect of ALE processing on Nb surface composition.  
Atomic concentrations extracted from XPS for both ALE and sputtered Nb surfaces are given in \Cref{tab:xps_atomic_concentrations}, and \Cref{fig:XPS} shows elemental spectra for Nb 3d, O 1s, F 1s, and C 1s. 
For clarity, each sample's spectra are normalized to their corresponding C 1s maximum, which is calibrated at a binding energy of 285.0 eV.
The Nb 3d spectra show that the ALE-treated niobium surface is predominantly in the \NbOx{} state (69\% fraction of the Nb 3d signal), with suboxides accounting for 8\% of the Nb signal.
In comparison, the sputtered surface shows a smaller amount of \NbOx{} (31\% of the total Nb signal) and slightly larger suboxide composition of 10\%.
The most significant distinction between the two samples is seen in the F 1s spectrum, which shows a distinct fluorine signal on the ALE-treated sample corresponding to an atomic fraction of 18\%.
These data support our expectations from ellipsometry that ALE forms a modified surface region in the Nb film containing both oxide and fluoride species.
It is unknown what effect this oxyfluoride passivation layer will have on applications of the ALE-treated films, and further investigations are needed to determine its effect on device performance.

\begin{table}[htbp]
    \centering
    \caption{Atomic concentrations from XPS.}
    \label{tab:xps_atomic_concentrations}
    \setlength{\tabcolsep}{4.5pt}
    \begin{tabular}{lcccccc}
        \toprule
        Sample & Nb (\%) & O (\%) & F (\%) & C (\%) & N (\%) & S (\%)\\
        \midrule
        Untreated & 30 & 48 & 1 & 16 & 5 & 0  \\
        ALE-treated & 16 & 43 & 18 & 14 & 6 & 3 \\
        \bottomrule
    \end{tabular}
\end{table}

\begin{figure*}
    \vspace{2 mm}
    \centering
    \includegraphics[width=0.75\linewidth]{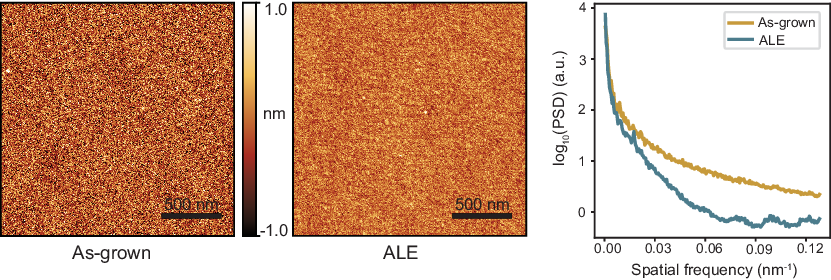}
    \caption{AFM scans of a 22.5 nm-thick Nb film as-grown by TLE (left) and after 5 cycles of ALE which decreases the thickness by 4 nm (center). The rms roughness $S_q$ decreases from  370 pm to 190 pm after ALE. (right) PSD of AFM scans for the as-grown (yellow lines) and ALE sample (blue line), highlighting the smoothing effect of ALE across all spatial frequencies.}
    \label{fig:AFM}
\end{figure*}

We next examined the effect of the ALE process on surface morphology. AFM scans of the as-grown TLE sample and ALE-treated samples are shown in \Cref{fig:AFM}. 
The as-grown TLE Nb film has a root-mean-square (rms) roughness ($S_q$) of $\sim$ 370 pm over a $2\times2~\mu m^2$ area. After etching 4 nm with ALE over 5 cycles, the roughness decreases to $\sim$ 190 pm over $2\times2~\mu m^2$. To examine the length-scale dependence of this smoothing, \Cref{fig:AFM} shows the two-dimensional isotropic power spectral density (PSD), calculated from the AFM topography using Bruker's NanoScope Analysis software.\cite{bruker_nanoscope_analysis_150} 
The PSD describes the distribution of surface roughness as a function of spatial frequency, allowing for surface features at different lateral length scales to be compared.
Following ALE, the PSD decreases across the spatial frequencies measured, indicating that the process smooths morphology over the lateral length scales characterized in the AFM scan.

\subsection{Cryogenic electrical and superconducting properties}
\begin{figure}
    \vspace{2 mm}
    \centering
    \includegraphics[width=1.0\linewidth]{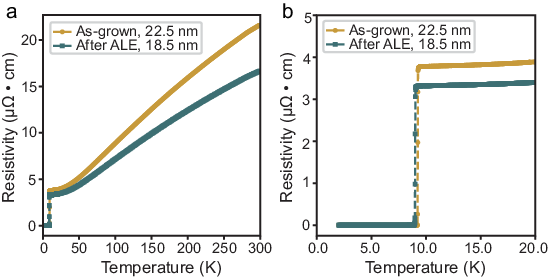}
    \caption{Resistivity versus temperature for TLE-grown Nb (yellow) and ALE-treated Nb (blue) cleaved from the same grown film. (a) Full resistivity drop from 300 K and (b) zoom-in around the critical temperature region. The $T_c$ of the as-grown is 9.24 K and 9.01 K after ALE. $\rho_{20 K}$ = 3.89 and 3.40 $\mu\Omega\cdot$cm for the as-grown and ALE samples, respectively. Overall, ALE does not significantly degrade the superconducting and electrical properties of the Nb thin film.}
    \label{fig:PPMS}
\end{figure}

Lastly, we evaluated the effect of ALE on the electrical and superconducting properties of the underlying Nb film.
\Cref{fig:PPMS} shows resistivity versus temperature curves for the TLE-grown Nb film that was etched from 22.5 to 18.5 nm by ALE and a pristine TLE-grown chip cleaved from the same starting sample.
\Cref{tab:ppms_properties} gives the extracted properties, including the $T_c$ (calculated from the midpoint of the transition), transition width ($\Delta T_c =T_{c,90}-T_{c,10}$), resistivity at 300 K and 20 K, and residual resistance ratio (RRR = $\rho_{300K}/\rho_{20K}$).

\begin{table}[htbp]
    \centering
    \caption{Cryogenic DC electrical properties.}
    \label{tab:ppms_properties}
    \setlength{\tabcolsep}{6.5pt}
    \begin{tabular}{lccccc}
        \toprule
        Sample & \makecell{$T_c$\\(K)} & \makecell{$\Delta T_c$\\(mK)} & \makecell{$\rho_{300K}$\\($\mu\Omega\cdot$cm)} & \makecell{$\rho_{20K}$\\($\mu\Omega\cdot$cm)} & RRR \\
        \midrule
        As-grown & 9.24 & 28.7 & 21.6 & 3.89 & 5.71  \\
        ALE-treated & 9.01 & 38.3 & 16.6 & 3.40 & 5.02  \\
        \bottomrule
    \end{tabular}
\end{table}

While the Nb film thickness is reduced by 15$\%$ from ALE, the $T_c$ is retained within $\sim$2.6$\%$ of the as-grown film. 
A reduction in $T_c$ is expected from the decreasing film thickness. \cite{pintoDimensional2018} 
Specifically, based on thickness dependence reported in ref.~\cite{pintoDimensional2018} a decrease in $T_c$ of $\sim$0.2--0.4 K is expected for a thickness change from 22.5--18.5 nm, which is compatible with our observations. 
We therefore conclude that ALE does not degrade $T_c$ beyond the decrease expected from reduced thickness.
Surprisingly, the room temperature resistivity is lower after ALE, which does not follow the expected increase normally associated with thinning or introduction of disorder.\cite{pintoDimensional2018, BianchiThicknessEffect2024a}
This behavior is therefore difficult to explain with previously established scaling theories for thickness, resistivity, and $T_c$.\cite{IvryUniversalScaling2014}
The RRR is slightly decreased and the superconducting transition width is broadened after ALE, however, suggesting that some disorder is introduced from the process; we expect that this effect could originate from the surface oxyfluoride layer.
Altogether, the modified surface composition does not significantly disrupt the DC electrical properties of the film, which is encouraging for the application of ALE for superconducting devices.

\section{Discussion}
We have reported an ALE process for Nb that simultaneously enables etching with Angstrom-scale precision while preserving the functional properties of the superconducting Nb film.
These capabilities are particularly relevant to superconducting devices where surface morphology and film dimensions impact performance.
The surface smoothing observed here arises from the self-limiting surface modification and removal chemistry of ALE, which can attenuate roughness and topographic features.\cite{gerritsenSurfaceSmoothing2022, kanarikAtomic2018}
Such smoothing may benefit niobium superconducting nanostructures, including SNSPD nanowires, where variations in nanowire geometry and line-edge roughness can influence critical current and device performance.\cite{Luo2023Niobium}
More broadly, the precise etching control provided by ALE enables precise adjustment of superconducting film thickness, which directly influences properties such as kinetic inductance\cite{gubinDependence2005} or critical current and can drive dimensional crossover and quantum size effects in sufficiently thin Nb films.\cite{pintoDimensional2018} 
Such thickness control could enable systematic access to different superconducting regimes, including the superconductor-insulator transition observed in ultrathin superconducting films.\cite{BottcherTransmonQubit2025}

The main open question for the ALE process demonstrated here is the functional impact of the oxyfluoride surface layer.
Our DC electrical data indicate that it does not disrupt $T_c$ or normal-state resistivity for $\sim$20 nm-thick Nb films, but assessment of effects on superconducting device characteristics such as microwave loss  or other surface-sensitive phenomena warrants additional investigation.
A correlation between greater fraction of Nb suboxides (relative to \NbOx{}) and reduced qubit $T_1$ relaxation times has been observed,\cite{PremkumarMicroscopicRelaxation2021} highlighting the potential significance of the reduced suboxide fractions produced by our ALE.
Furthermore, plasma passivation and modification of the near-surface Nb region with nitridation has been shown to reduce microwave loss,\cite{ZhengNitrogenPlasma2022} and it is possible that fluorination could accomplish a similar effect in our case.
Together, these results suggest that the oxyfluoride surface layer may influence superconducting device performance, although its specific functional role remains to be determined.

Lastly, further exploration of the ALE process itself could examine how process conditions influence surface modification and feature fidelity.
Although tool limitations prevented us from investigating temperatures $<{125}~^\circ\text{C}$, lower process temperature may reduce thermal diffusion and decrease the thickness of the oxyfluoride surface layer.
Control of the surface modification depth could also affect the soft saturation behavior observed with the \HF{} half step and reduce etching beyond the intentionally modified region.
Further work could also examine the anisotropy of the process and its ability to preserve feature fidelity in device-relevant geometries.

\section{Conclusion}
We have demonstrated an ALE process for Nb that enables etching with sub-nanometer precision and surface smoothing without degrading the DC superconducting or electrical properties of the material.
The cyclic O$_2$ plasma modification and \HF{} removal process enables precise material removal while reducing surface roughness.
Despite the presence of a remaining oxyfluoride surface layer, the superconducting transition temperature is not degraded beyond the expected decrease due to reduced film thickness after ALE. 
Further investigation of the functional consequences of the oxyfluoride surface layer, particularly for surface-sensitive superconducting device properties, will be important for understanding its role in device performance. 
Overall, this work provides a foundation for using ALE to precisely control Nb thickness and surface morphology while retaining its superconducting properties.

\section*{Acknowledgments}
L.H.W. acknowledges support from the IQIM/AWS Postdoctoral Fellowship and the Broadcom Innovation Fund. A. A. H. was supported by the National Science Foundation under Award \#2234390. 
Y. Y. was supported by a Resnick Explorer grant. A. J. M. was supported by the Broadcom Innovation Fund. We gratefully acknowledge the critical support and infrastructure provided for this work by The Kavli Nanoscience Institute and the Molecular Materials Research Center of the Beckman Institute at the California Institute of Technology.

\section*{Author declarations}
The authors have no conflicts to disclose.

\section*{Author contributions}
Lillian B. Hughes Wyatt: Conceptualization (equal); Data curation (equal); Formal analysis (equal); Investigation (equal); Methodology (equal); Validation (equal); Visualization (equal); Writing – original draft (equal); Writing – review \& editing (equal). Azmain A. Hossain: Conceptualization (supporting); Writing – review \& editing (supporting). Yifei Yan: Investigation (supporting). Austin J. Minnich: Conceptualization (equal); Formal analysis (equal); Funding acquisition (equal); Methodology (equal); Project administration (equal); Resources (equal); Supervision (equal); Validation (equal); Visualization (equal); Writing – review \& editing (equal).

\section*{Data Availability}
The data that support the findings of this study are available from the corresponding author upon reasonable request.

\section*{References}

\bibliography{NbALE}

@PREAMBLE{
 "\providecommand{\noopsort}[1]{}" 
 # "\providecommand{\singleletter}[1]{#1}%" 
}

@article{hossainAtomic2025,
    title = {Atomic layer etching of niobium nitride using sequential exposures of {O2} and {H2}/{SF6} plasmas},
    volume = {43},
    issn = {0734-2101, 1520-8559},
    url = {https://pubs.aip.org/jva/article/43/4/042605/3351654/Atomic-layer-etching-of-niobium-nitride-using},
    doi = {10.1116/6.0004548},
    number = {4},
    urldate = {2025-11-04},
    journal = {Journal of Vacuum Science \& Technology A},
    author = {Hossain, Azmain A. and Murphy, Sela and Catherall, David S. and Ardizzi, Anthony J. and Minnich, Austin J.},
    month = jul,
    year = {2025},
    pages = {042605},
}

@misc{NISTXPS,
    author       = {{National Institute of Standards and Technology}},
    title        = {NIST X-ray Photoelectron Spectroscopy Database},
    year         = {2023},
    howpublished = {NIST Standard Reference Database 20, Version 5.0},
    address      = {Gaithersburg, MD},
    doi          = {10.18434/T4T88K},
    url          = {https://srdata.nist.gov/xps/}
}

@article{hossainIsotropic2023,
    title = {Isotropic plasma-thermal atomic layer etching of superconducting titanium nitride films using sequential exposures of molecular oxygen and {SF6}/{H2} plasma},
    volume = {41},
    issn = {0734-2101, 1520-8559},
    url = {https://pubs.aip.org/jva/article/41/6/062601/2912772/Isotropic-plasma-thermal-atomic-layer-etching-of},
    doi = {10.1116/6.0002965},
    number = {6},
    urldate = {2025-11-21},
    journal = {Journal of Vacuum Science \& Technology A},
    author = {Hossain, Azmain A. and Wang, Haozhe and Catherall, David S. and Leung, Martin and Knoops, Harm C. M. and Renzas, James R. and Minnich, Austin J.},
    month = dec,
    year = {2023},
    pages = {062601},
}

@article{Dhas2025,
    author = {Dhas, Jeffrey
A. and Bhatia, Ekta and Koirala, Krishna P. and Zhu, Zihua and Liu, Mingzhao and Schujman, Sandra and Rebar, Drew J. and Ponce, Francisco and Warner, Marvin G. and Chang, Chih-Hung and Murray, Thomas and Kirkpatrick, Sean and Sushko, Peter V. and Papa Rao, Satyavolu S. and Du, Yingge},
    title = {Tailoring
the Physicochemical Properties of Nb Thin
Films via Surface Engineering Methods},
    journal = {ACS Applied Materials \& Interfaces},
    volume = {17},
    number = {16},
    pages = {24502-24512},
    year = {2025},
    month = {04},
    issn = {1944-8244},
    doi = {10.1021/acsami.4c20490},
    url = {https://doi.org/10.1021/acsami.4c20490},
}

@article{Kar2023,
    author = {Kar, Soumen and Weiland, Conan and Zhou, Chenyu and Bhatia, Ekta and Martinick, Brian and Nalaskowski, Jakub and Mucci, John and Olson, Stephen and Hung, Pui Yee and Wells, Ilyssa and Frost, Hunter and Johnson, Corbet S. and Murray, Thomas and Kaushik, Vidya and Kirkpatrick, Sean and Chau, Kiet and Walsh, Michael J. and Liu, Mingzhao and Papa Rao, Satyavolu S.},
    title = {Engineering of niobium surfaces through accelerated neutral atom beam technology for quantum applications},
    journal = {Journal of Applied Physics},
    volume = {134},
    number = {2},
    pages = {025301},
    year = {2023},
    month = {07},
    issn = {0021-8979},
    doi = {10.1063/5.0153617},
    url = {https://doi.org/10.1063/5.0153617}
}

@article{Prudnikava2022Niobium,
    author  = {Prudnikava, A. and Tamashevich, Y. and Babenkov, S. and Makarova, A. and Smirnov, D. and Aristov, V. and Molodtsova, O. and Kugeler, O. and Viefhaus, J. and Foster, B.},
    title   = {Systematic study of niobium thermal treatments for superconducting radio frequency cavities employing x-ray photoelectron spectroscopy},
    journal = {Superconductor Science and Technology},
    volume  = {35},
    number  = {6},
    pages   = {065019},
    year    = {2022},
    doi     = {10.1088/1361-6668/ac6a85}
}

@misc{Yifei,
    title = {Thermal laser evaporation of superconducting Nb films},
    doi = {arXiv:2610.38070},
    publisher = {arXiv},
    author = {Yifei Yan and David S. Catherall and Finley B. Donachie and Dorsa Najafipanah and Austin J. Minnich},
    year = {2026}
}

@manual{bruker_nanoscope_analysis_150,
  author       = {{Bruker}},
  title        = {{NanoScope Analysis 1.50 User Manual}},
  organization = {Bruker},
  url          = {https://www.epfl.ch/research/facilities/cmi/wp-content/uploads/2020/03/NanoScopeAnalysisV150Manual.pdf},
  note         = {Accessed: 2026-09-11}
}

@article{pintoDimensional2018,
    title = {Dimensional crossover and incipient quantum size effects in superconducting niobium nanofilms},
    volume = {8},
    issn = {2045-2322},
    url = {https://www.nature.com/articles/s41598-018-22983-6},
    doi = {10.1038/s41598-018-22983-6},
    number = {1},
    urldate = {2025-12-03},
    journal = {Scientific Reports},
    author = {Pinto, Nicola and Rezvani, S. Javad and Perali, Andrea and Flammia, Luca and Milošević, Milorad V. and Fretto, Matteo and Cassiago, Cristina and De Leo, Natascia},
    month = mar,
    year = {2018},
    pages = {4710},
}

@article{BianchiThicknessEffect2024a,
    title = {Thickness effect on superconducting properties of niobium films for radio-frequency cavity applications},
    volume = {37},
    issn = {0953-2048, 1361-6668},
    url = {https://iopscience.iop.org/article/10.1088/1361-6668/ad59ce},
    doi = {10.1088/1361-6668/ad59ce},
    number = {8},
    urldate = {2026-09-11},
    journal = {Superconductor Science and Technology},
    author = {Bianchi, Antonio and Bonura, Marco and Carlos, Carlota P A and Leith, Stewart and Rosaz, Guillaume and Senatore, Carmine and Venturini Delsolaro, Walter},
    month = aug,
    year = {2024},
    pages = {085005},
}

@article{IvryUniversalScaling2014,
    title = {Universal scaling of the critical temperature for thin films near the superconducting-to-insulating transition},
    volume = {90},
    copyright = {http://creativecommons.org/licenses/by/3.0/},
    issn = {1098-0121, 1550-235X},
    url = {https://link.aps.org/doi/10.1103/PhysRevB.90.214515},
    doi = {10.1103/PhysRevB.90.214515},
    number = {21},
    urldate = {2026-09-08},
    journal = {Physical Review B},
    author = {Ivry, Yachin and Kim, Chung-Soo and Dane, Andrew E. and De Fazio, Domenico and McCaughan, Adam N. and Sunter, Kristen A. and Zhao, Qingyuan and Berggren, Karl K.},
    month = dec,
    year = {2014},
    pages = {214515},
}

@article{GiacconeStudyNiobium2022,
    title = {Study of the {Niobium} {Oxide} {Structure} and {Microscopic} {Effect} of {Plasma} {Processing} on the {Niobium} {Surface}},
    volume = {SRF2021},
    copyright = {Creative Commons Attribution 4.0 International},
    issn = {2673-5504},
    url = {https://jacow.org/srf2021/doi/JACoW-SRF2021-WEPCAV001.html},
    doi = {10.18429/JACOW-SRF2021-WEPCAV001},
    urldate = {2026-09-14},
    journal = {Proceedings of the 20th International Conference on RF Superconductivity},
    publisher = {JACoW Publishing, Geneva, Switzerland},
    author = {Giaccone, Bianca and Martinello, Martina and Zasadzinski, John},
    editor = {Kenji (Ed.), Saito and Ting (Ed.), Xu and Naruhiko (Ed.), Sakamoto and Ana (Ed.), Lesage and R.W. (Ed.), Volker, Schaa},
    year = {2022},
    pages = {5 pages, 1.846 MB},
}

@article{Luo2023Niobium,
  author  = {Luo, Peng and Zhao, Yihui},
  title   = {Niobium Nitride Preparation for Superconducting Single-Photon Detectors},
  journal = {Molecules},
  volume  = {28},
  number  = {17},
  pages   = {6200},
  year    = {2023},
  doi     = {10.3390/molecules28176200}
}

@misc{BottcherTransmonQubit2025,
    title = {A transmon qubit realized by exploiting the superconductor-insulator transition},
    doi = {10.48550/arXiv.2510.19983},
    publisher = {arXiv},
    author = {Bøttcher, C. G. L. and Önder, E. and Connolly, T. and Zhao, J. and Kvande, C. and Wang, D. Q. and Kurilovich, P. D. and Vaitiekėnas, S. and Glazman, L. I. and Tang, H. X. and Devoret, M. H.},
    year = {2025},
    note = {arXiv:2510.19983 [quant-ph]},
}

@article{ZhengNitrogenPlasma2022,
  title = {Nitrogen plasma passivated niobium resonators for superconducting quantum circuits},
  volume = {120},
  issn = {0003-6951, 1077-3118},
  url = {https://pubs.aip.org/apl/article/120/10/102601/2833153/Nitrogen-plasma-passivated-niobium-resonators-for},
  doi = {10.1063/5.0082755},
  number = {10},
  urldate = {2026-10-26},
  journal = {Applied Physics Letters},
  author = {Zheng, K. and Kowsari, D. and Thobaben, N. J. and Du, X. and Song, X. and Ran, S. and Henriksen, E. A. and Wisbey, D. S. and Murch, K. W.},
  month = mar,
  year = {2022},
  pages = {102601},
}

@article{PremkumarMicroscopicRelaxation2021,
  title = {Microscopic relaxation channels in materials for superconducting qubits},
  volume = {2},
  issn = {2662-4443},
  url = {https://www.nature.com/articles/s43246-021-00174-7},
  doi = {10.1038/s43246-021-00174-7},
  number = {1},
  urldate = {2025-12-03},
  journal = {Communications Materials},
  author = {Premkumar, Anjali and Weiland, Conan and Hwang, Sooyeon and Jäck, Berthold and Place, Alexander P. M. and Waluyo, Iradwikanari and Hunt, Adrian and Bisogni, Valentina and Pelliciari, Jonathan and Barbour, Andi and Miller, Mike S. and Russo, Paola and Camino, Fernando and Kisslinger, Kim and Tong, Xiao and Hybertsen, Mark S. and Houck, Andrew A. and Jarrige, Ignace},
  month = jul,
  year = {2021},
  pages = {72},
}

@article{kanarikOverview2015,
  title = {Overview of atomic layer etching in the semiconductor industry},
  volume = {33},
  issn = {0734-2101, 1520-8559},
  url = {https://pubs.aip.org/jva/article/33/2/020802/246821/Overview-of-atomic-layer-etching-in-the},
  doi = {10.1116/1.4913379},
  number = {2},
  urldate = {2025-11-25},
  journal = {Journal of Vacuum Science \& Technology A: Vacuum, Surfaces, and Films},
  author = {Kanarik, Keren J. and Lill, Thorsten and Hudson, Eric A. and Sriraman, Saravanapriyan and Tan, Samantha and Marks, Jeffrey and Vahedi, Vahid and Gottscho, Richard A.},
  month = mar,
  year = {2015},
  pages = {020802},
}

@article{kanarikAtomic2018,
  title = {Atomic {Layer} {Etching}: {Rethinking} the {Art} of {Etch}},
  volume = {9},
  copyright = {http://pubs.acs.org/page/policy/authorchoice\_termsofuse.html},
  issn = {1948-7185, 1948-7185},
  shorttitle = {Atomic {Layer} {Etching}},
  url = {https://pubs.acs.org/doi/10.1021/acs.jpclett.8b00997},
  doi = {10.1021/acs.jpclett.8b00997},
  number = {16},
  urldate = {2025-11-24},
  journal = {The Journal of Physical Chemistry Letters},
  author = {Kanarik, Keren J. and Tan, Samantha and Gottscho, Richard A.},
  month = aug,
  year = {2018},
  pages = {4814--4821},
}

@article{WalshAtomicLayer2026a,
  title = {Atomic layer etch process for {Nb} and {Ta} using {CF4}/{H2} plasma},
  volume = {44},
  issn = {0734-2101, 1520-8559},
  url = {https://pubs.aip.org/jva/article/44/2/020403/3378527/Atomic-layer-etch-process-for-Nb-and-Ta-using-CF4},
  doi = {10.1116/6.0005119},
  number = {2},
  urldate = {2026-03-12},
  journal = {Journal of Vacuum Science \& Technology A},
  author = {Walsh, Ryan F. and Macera, Mikeal B. and Renzas, J. Russell},
  month = mar,
  year = {2026},
  pages = {020403},
}

@article{Pankratiev2020Etching,
    author = {Pankratiev, P. A. and Barsukov, Yu. V. and Kobelev, A. A. and Vinogradov, A. Ya. and Miroshnikov, I. V. and Smirnov, A. S.},
    title = {Etching of Si$_3$N$_4$ by SF$_6$/H$_2$ and SF$_6$/D$_2$ plasmas},
    journal = {Journal of Physics: Conference Series},
    volume = {1697},
    number = {1},
    pages = {012222},
    year = {2020},
    doi = {10.1088/1742-6596/1697/1/012222}
}

@article{Jung2020Highly,
    author = {Jung, Ji-Eun and Barsukov, Yuri and Volynets, Vladimir and Kim, Gonjun and Nam, Sang Ki and Han, Kyuhee and Huang, Shuo and Kushner, Mark J.},
    title = {Highly selective Si$_3$N$_4$/SiO$_2$ etching using an NF$_3$/N$_2$/O$_2$/H$_2$ remote plasma. II. Surface reaction mechanism},
    journal = {Journal of Vacuum Science \& Technology A},
    volume = {38},
    number = {2},
    pages = {023008},
    year = {2020},
    doi = {10.1116/1.5125569}
}

@article{mahuliImproving2025,
  title = {Improving the {Lifetime} of {Aluminum}-{Based} {Superconducting} {Qubits} through {Atomic} {Layer} {Etching} and {Deposition}},
  copyright = {https://doi.org/10.15223/policy-029},
  issn = {1936-0851, 1936-086X},
  url = {https://pubs.acs.org/doi/10.1021/acsnano.5c14022},
  doi = {10.1021/acsnano.5c14022},
  urldate = {2025-11-24},
  journal = {ACS Nano},
  author = {Mahuli, Neha and Minguzzi, Joaquin and Gao, Jiansong and Resnick, Rachel and Diez, Sandra and Cosmic, R. and Marcaud, Guillaume and Hunt, Matthew and Swenson, Loren and Rose, Jefferson and Painter, Oskar and Jarrige, Ignace},
  month = nov,
  year = {2025},
  pages = {acsnano.5c14022},
}

@article{Gurevich2017Theory,
  author = {Gurevich, A.},
  title = {Theory of RF superconductivity for resonant cavities},
  journal = {Superconductor Science and Technology},
  volume = {30},
  number = {3},
  pages = {034004},
  year = {2017},
  doi = {10.1088/1361-6668/30/3/034004}
}

@article{Zmuidzinas2012Superconducting,
  author = {Zmuidzinas, Jonas},
  title = {Superconducting Microresonators: Physics and Applications},
  journal = {Annual Review of Condensed Matter Physics},
  volume = {3},
  pages = {169--214},
  year = {2012},
  doi = {10.1146/annurev-conmatphys-020911-125022}
}

@article{SiddiqiEngineeringHighcoherence2021c,
  title = {Engineering high-coherence superconducting qubits},
  volume = {6},
  issn = {2058-8437},
  url = {https://www.nature.com/articles/s41578-021-00370-4},
  doi = {10.1038/s41578-021-00370-4},
  number = {10},
  urldate = {2026-02-14},
  journal = {Nature Reviews Materials},
  author = {Siddiqi, Irfan},
  month = sep,
  year = {2021},
  pages = {875--891},
}

@article{Semenov2003Ultimate,
  author = {Semenov, A. and others},
  title = {Ultimate performance of a superconducting quantum detector},
  journal = {The European Physical Journal Applied Physics},
  volume = {21},
  number = {3},
  pages = {171--178},
  year = {2003},
  doi = {10.1051/epjap:2003009}
}

@article{murrayMaterial2021,
  title = {Material matters in superconducting qubits},
  volume = {146},
  issn = {0927796X},
  url = {https://linkinghub.elsevier.com/retrieve/pii/S0927796X21000413},
  doi = {10.1016/j.mser.2021.100646},
  urldate = {2025-11-24},
  journal = {Materials Science and Engineering: R: Reports},
  author = {Murray, Conal E.},
  month = oct,
  year = {2021},
  pages = {100646},
}

@article{gubinDependence2005,
  title     = {Dependence of magnetic penetration depth on the thickness of superconducting {Nb} thin films},
  author    = {Gubin, A. I. and Il'in, K. S. and Vitusevich, S. A. and Siegel, M. and Klein, N.},
  journal   = {Physical Review B},
  volume    = {72},
  number    = {6},
  pages     = {064503},
  year      = {2005},
  publisher = {American Physical Society},
  doi       = {10.1103/PhysRevB.72.064503}
}

@article{anferovImproved2024,
  title = {Improved coherence in optically defined niobium trilayer-junction qubits},
  volume = {21},
  issn = {2331-7019},
  url = {https://link.aps.org/doi/10.1103/PhysRevApplied.21.024047},
  doi = {10.1103/PhysRevApplied.21.024047},
  number = {2},
  urldate = {2025-11-22},
  journal = {Physical Review Applied},
  author = {Anferov, Alexander and Lee, Kan-Heng and Zhao, Fang and Simon, Jonathan and Schuster, David I.},
  month = feb,
  year = {2024},
  pages = {024047},
}

@article{wallraffStrong2004,
  author  = {Wallraff, Andreas and Schuster, David I. and Blais, Alexandre and
             Frunzio, Luigi and Huang, R.-S. and Majer, Johannes and
             Kumar, Sameer and Girvin, Steven M. and Schoelkopf, Robert J.},
  title   = {Strong coupling of a single photon to a superconducting qubit using circuit quantum electrodynamics},
  journal = {Nature},
  volume  = {431},
  number  = {7005},
  pages   = {162--167},
  year    = {2004},
  doi     = {10.1038/nature02851}
}

@article{altoeLocalization2022a,
  title = {Localization and {Mitigation} of {Loss} in {Niobium} {Superconducting} {Circuits}},
  volume = {3},
  issn = {2691-3399},
  url = {https://link.aps.org/doi/10.1103/PRXQuantum.3.020312},
  doi = {10.1103/PRXQuantum.3.020312},
  number = {2},
  urldate = {2025-12-08},
  journal = {PRX Quantum},
  author = {Altoé, M. Virginia P. and Banerjee, Archan and Berk, Cassidy and Hajr, Ahmed and Schwartzberg, Adam and Song, Chengyu and Alghadeer, Mohammed and Aloni, Shaul and Elowson, Michael J. and Kreikebaum, John Mark and Wong, Ed K. and Griffin, Sinéad M. and Rao, Saleem and Weber-Bargioni, Alexander and Minor, Andrew M. and Santiago, David I. and Cabrini, Stefano and Siddiqi, Irfan and Ogletree, D. Frank},
  month = apr,
  year = {2022},
  pages = {020312},
}

@article{deLeonMaterials2021,
  author  = {de Leon, Nathalie P. and Itoh, Kohei M. and Kim, Dohun and
             Mehta, Karan K. and Northup, Tracy E. and Paik, Hanhee and
             Palmer, B. S. and Samarth, N. and Sangtawesin, Sorawis and
             Steuerman, D. W.},
  title   = {Materials challenges and opportunities for quantum computing hardware},
  journal = {Science},
  volume  = {372},
  number  = {6539},
  pages   = {eabb2823},
  year    = {2021},
  doi     = {10.1126/science.abb2823}
}

@article{lichtenbergerInvestigation1993,
  author  = {Lichtenberger, A. W. and Lea, D. M. and Lloyd, F. L.},
  title   = {Investigation of etching techniques for superconductive {Nb}/{Al}-{Al2O3}/{Nb} fabrication processes},
  journal = {IEEE Transactions on Applied Superconductivity},
  volume  = {3},
  number  = {1},
  pages   = {2191--2196},
  year    = {1993},
  doi     = {10.1109/77.233938}
}

@article{drimmerEffect2025,
  author  = {Drimmer, Maxwell and Telkamp, Sjoerd and
             Fischer, Felix L. and Rodrigues, Ines C. and
             Todt, Clemens and Krizek, Filip and
             Kriegner, Dominik and M{\"u}ller, Christoph and
             Wegscheider, Werner and Chu, Yiwen},
  title   = {The effect of niobium thin film structure on
             losses in superconducting circuits},
  journal = {Materials for Quantum Technology},
  volume  = {5},
  number  = {4},
  pages   = {041001},
  year    = {2025},
  doi     = {10.1088/2633-4356/ae1347}
}

@article{krieger2025atomic,
  author = {Krieger, Guillaume and
            Peeters, Silke A. and
            Vonken, Bob J. M. and
            Chittock, Nicholas J. and
            Mackus, Adriaan J. M. and
            Kessels, Wilhelmus M. M. and
            Knoops, Harm C. M.},
  title = {Atomic layer etching of {TiN} by oxidation and
           {SF6-H2} plasma exposure: {ALE} window defined
           by {HF}, {H}, and {F}},
  journal = {Applied Physics Letters},
  volume = {127},
  number = {19},
  pages = {191606},
  year = {2025},
  month = nov,
  doi = {10.1063/5.0285942}
}

@article{gerritsenSurfaceSmoothing2022,
  author = {Gerritsen, Sven H. and
            Chittock, Nicholas J. and
            Vandalon, Vincent and
            Verheijen, Marcel A. and
            Knoops, Harm C. M. and
            Kessels, Wilhelmus M. M. and
            Mackus, Adriaan J. M.},
  title = {Surface Smoothing by Atomic Layer Deposition and Etching
           for the Fabrication of Nanodevices},
  journal = {ACS Applied Nano Materials},
  volume = {5},
  number = {12},
  pages = {18116--18126},
  year = {2022},
  month = dec,
  doi = {10.1021/acsanm.2c04025}
}

\end{document}